\documentclass[aps,twocolumn,prb,showpacs]{revtex4}

\usepackage[dvips]{color}
\usepackage{graphicx}
\usepackage{epsfig}
\usepackage{dcolumn}
\usepackage{color}
\usepackage{amssymb}
\usepackage[english]{babel}
\usepackage{mathrsfs}
\usepackage{bm}
\usepackage{amsmath}
\usepackage{lscape}

\begin{document}

\title{ Single-photon nonlinear optics with Kerr-type nanostructured materials}

\author{Sara Ferretti and Dario Gerace$^{*}$}
\affiliation{Dipartimento di Fisica ``Alessandro Volta'',
Universit\`a di Pavia,  via Bassi 6, I-27100 Pavia, Italy}


\date{\today}

\begin{abstract}
We employ a quantum theory of the nonlinear optical response from 
an actual solid-state material possessing an intrinsic bulk contribution to the 
third-order nonlinear susceptibility (Kerr-type nonlinearity), which can be 
arbitrarily nanostructured to achieve diffraction-limited electromagnetic 
field confinement. 
By calculating the zero-time delay second-order correlation of the cavity 
field, we set the conditions for using semiconductor or insulating materials with 
near-infrared energy gaps 
as efficient means to obtain single-photon nonlinear behavior in prospective 
solid-state integrated devices, alternative to ideal sources 
of quantum radiation such as, e.g., single two-level emitters. 
\end{abstract}

\maketitle

Quantum information processing based on photonic platforms is one of the most
promising routes towards a fully integrated technology exploiting the laws 
of quantum mechanics.\cite{qp_review} In this context, many quantum optical tasks, 
such as single-photon switches and two-qubit quantum gates, would require strong 
photon-photon interactions - ultimately at the single-photon level - to be engineered
in solid state devices.\cite{imamoglu99prl} 
Besides being of practical interest for prospective applications in quantum photonics, 
strongly correlated photonic systems promise fascinating perspectives for a number of 
theoretical proposals concerning the manybody behavior of complex 
nonlinear and tunnel-coupled devices.\cite{hartmann06,gerace_josephson,ferretti2010}

Cavity quantum electrodynamics (CQED) is the most straightforward way of obtaining 
single-photon nonlinear behavior, thanks to the underlying anharmonicity introduced 
by a single atomic-like emitter into a high-finesse resonator.\cite{tian92,imamoglu99}
It has been shown experimentally with single caesium atoms strongly coupled to a Fabry-P\'erot 
resonant mode\cite{birnbaum05nat} that such a system is able to block the transmission 
of a single photon when another photon is present in the cavity:  
a \textit{photon blockade} effect.\cite{imamoglu97}
The quantum efficiency of this process is operationally determined by the 
degree of antibunching in the second-order correlation function  
for the emitted radiation, after resonant excitation of the CQED system.\cite{rebic04pra} 
Analogous effects have been measured in solid-state systems with quantum dots coupled 
to dielectric resonators, both under nonresonant\cite{kevin07nat} and resonant\cite{faraon08nphys} 
excitation conditions.

\begin{figure}[b]
\begin{center}
\vspace{-0.5cm}
\includegraphics[width=0.48\textwidth]{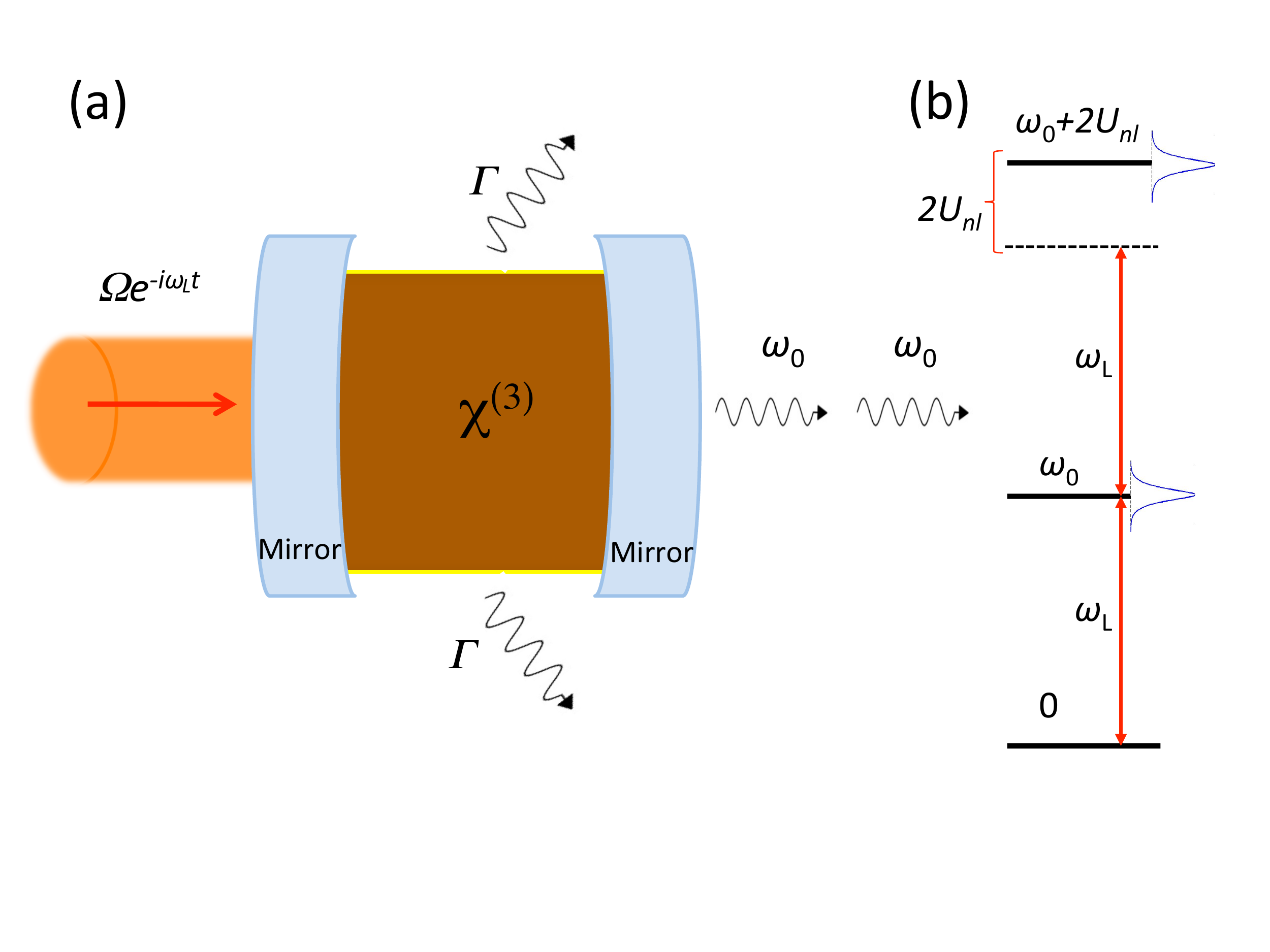}
\vspace{-1.5cm} 
\caption{(Color online) (a) Scheme of a resonator made of a Kerr-type 
   nonlinear material, which is resonantly driven by a coherent field and undergoes
   single-photon blockade, and (b) energy level diagram giving rise to the emission 
   of single photon Fock states from the cavity.} \label{fig1}
\end{center}
\end{figure}

The photon blockade can be realized when two photons inside a resonant system
produce a nonlinear shift of its response frequency, $U_{\mathrm{nl}}$, that is 
larger than the line-broadening induced by losses and decoherence rate, $\Gamma$, 
as shown in Fig.~\ref{fig1}.
Theoretical proposals to achieve single-photon nonlinearities in solid-state 
systems usually rely on enhanced light-matter coupling of some dipole-allowed transition, 
where material excitations can provide the required quantum anharmonicity. 
Strong Kerr-type nonlinearities are predicted for single atomic-like transitions 
coupled to high-quality resonators,\cite{auffeves07,jacobs09prl} or in
strongly confined polaritonic systems.\cite{ciuti06prb,carusot09preprint}
For semiconductor microcavities, strong coupling of single photons 
mediated by enhanced second-order nonlinearity [$\chi^{(2)}$] has been 
theoretically discussed.\cite{irvine06prl}
It has also been predicted that  suitably engineered coupled cavities can 
considerably relax the requirement on the condition that the effective nonlinear 
interaction be larger than the fundamental resonance linewidth.\cite{savona10prl,bamba}
However, owing to the intrinsically small value of the third-order nonlinear susceptibilities
[$\chi^{(3)}$] in ordinary bulk media,\cite{boyd_book} it is commonly accepted 
that appreciable resonance shifts for nonlinear materials in their transparency 
optical range would require an exponentially large number of photons. 

In this work, we challenge the latter idea by quantitatively showing that a realistic
nanostructuring of an ordinary nonlinear medium is able to produce very large
effective nonlinear susceptibilities, ultimately sensitive at the single-photon level. 
From a canonical quantization of the classical nonlinear optical response for a 
single mode of the electromagnetic field, we solve the quantum master equation for 
the system density matrix, where the real part of $\chi^{(3)}$ is related to the effective 
photon-photon interaction energy\cite{loudon_book} and losses of the resonator 
mode, such as coupling to free space modes or two-photon absorption, are fully 
taken into account. 

Throughout this work, we adopt the classical nonlinear 
optics notation in SI units.\cite{boyd_book}
The nonlinear optical response to the applied electric field
of a generic dielectric material is given by 
\begin{eqnarray}\label{eq:induction}
D_i(\mathbf{r},t) &=& \varepsilon_{0} \varepsilon_{ij}(\mathbf{r}) E_{j}(\mathbf{r},t) + 
\varepsilon_{0}[ \chi^{(2)}_{ijk}(\mathbf{r})E_{j}(\mathbf{r},t)E_{k}(\mathbf{r},t)  \nonumber \\
&+&\chi^{(3)}_{ijkl}(\mathbf{r})E_{j}(\mathbf{r},t)E_{k}(\mathbf{r},t)E_{l}(\mathbf{r},t)+...]  \, ,  
\end{eqnarray}
where we employ the usual sum rule over repeated indices labeling the three spatial coordinates.
This relation defines the relative dielectric permettivity tensor of the medium,
$\varepsilon_{ij}(\mathbf{r})= \delta_{ij} + \chi^{(1)}_{ij}(\mathbf{r}) $.
We will now specify the nonlinear response to the case of a single
mode of the electromagnetic field inside a centrosymmetric medium, i.e. we
assume $\chi^{(2)}_{ijk}(\mathbf{r})=0$,\cite{boyd_book} and only consider 
Kerr-type nonlinear effects due to the $\chi^{(3)}$ tensor elements in Eq.~(\ref{eq:induction}). 
We assume an isotropic medium, i.e. a spatially dependent but scalar dielectric response, 
$\varepsilon_{ij}(\mathbf{r}) \to \varepsilon(\mathbf{r})$. 
The canonical quantization of a single mode of the electromagnetic field 
in a generic spatially dependent nonlinear medium is obtained after expressing
the quantized field operators for a single cavity mode as 
\begin{equation}\label{fields}
\hat{\mathbf{E}}(\mathbf{r},t) = i \left(\frac{\hbar\omega_0}{2\epsilon_{0}}\right)^{1/2}
\left[\hat{a}\frac{{\vec{\alpha}}(\mathbf{r})}{\sqrt{\varepsilon(\mathbf{r})}} e^{-i \omega_0 t}-
\hat{a}^{\dagger} \frac{{\vec{\alpha}}^{\ast}(\mathbf{r})}{\sqrt{\varepsilon(\mathbf{r})}} e^{i \omega_0 t}
\right]       \, ,     
\end{equation}
and $\hat{\mathbf{B}}(\mathbf{r})=(-i/ \omega_0)\nabla\times \hat{\mathbf{E}}(\mathbf{r})$,
where $\hat{a}$ ($\hat{a}^{\dagger}$) defines the destruction (creation) operator
of a single photon in the mode, and $\vec{\alpha}(\mathbf{r})$ is the normalized
three-dimensional cavity field profile satisfying the condition 
$\int |\vec{\alpha}(\mathbf{r})|^2 \mathrm{d}\mathbf{r} =1$.
From the classical expression of the time-averaged total energy density in the mode,
$\mathcal{H}_{em}= 
\frac{1}{2}\int\left[\mathbf{E}(\mathbf{r})\cdot\mathbf{D}(\mathbf{r})
+ \mathbf{H}(\mathbf{r})\cdot\mathbf{B}(\mathbf{r}) \right] \mathrm{d}\mathbf{r} \, $ 
(assuming $\mathbf{H}=\mathbf{B}/ \mu_0$ in a non-magnetic medium),
a nonlinear second-quantized hamiltonian can be eventually obtained  
\begin{equation}
\hat{H}=\hbar\omega_0 \hat{a}^{\dagger}\hat{a}  
+ \hat{H}_{\mathrm{nl}} \, .
\end{equation}
The linear part is the expected hamiltonian of a single harmonic oscillator (neglecting 
the zero point energy). In the nonlinear part we only retain the Kerr-type terms,\cite{sipe2004pre}
$\hat{H}_{\mathrm{nl}}=U_{\mathrm{nl}} \hat{a}^{\dagger}\hat{a}^{\dagger}\hat{a}\hat{a} $ with the
photon-photon interaction given by
\begin{equation}\label{ham_kerr}
U_{\mathrm{nl}}= \frac{D(\hbar\omega_0)^2 } {8\varepsilon_0} 
\int \mathrm{d}\mathbf{r} \,\, {\alpha}^{\ast}_i(\mathbf{r}) \frac{\mathrm{Re}\{\chi^{(3)}_{ijkl}(\mathbf{r})\}}{\varepsilon^2(\mathbf{r})} 
{\alpha}^{\ast}_j(\mathbf{r}){\alpha}_k(\mathbf{r}){\alpha}_l(\mathbf{r})  \, ,
\end{equation}
with degeneracy $D=6$. 
Equation (\ref{ham_kerr})  is a general expression for the nonlinear shift induced by the 
Kerr-effect at the single photon level, for an arbitrary spatially dependent response, 
such as photonic  crystal or pillar microcavities with 
ordinary nonlinear semiconductor materials.\cite{vahala_review} 
Even if the full expression should be applied to the specific case of interest for 
given nonlinear tensor components, we can simplify Eq.~(\ref{ham_kerr}) to give 
some quantitative estimates\cite{drummond80}
\begin{equation}\label{shift_kerr}
U_{\mathrm{nl}}\simeq \frac{3(\hbar\omega_0)^2 } {4\varepsilon_0}
\frac{\overline{\chi}^{(3)}}{\overline{\varepsilon}_r^2}
\int  |\vec{\alpha}(\mathbf{r})|^4 \mathrm{d}\mathbf{r} 
=  \frac{3(\hbar\omega_0)^2 } {4\varepsilon_0 V_{\mathrm{eff}}}
\frac{\overline{\chi}^{(3)}}{\overline{\varepsilon}_r^2} \, ,
\end{equation}
where the effective cavity mode volume is defined as 
$V^{-1}_{\mathrm{eff}}=\int  |\vec{\alpha}(\mathbf{r})|^4 \mathrm{d}\mathbf{r}$
within our formalism.
To have order of magnitude results, we assume constant values for the average 
real part of the nonlinear susceptibility and relative dielectric permittivity, $\overline{\chi}^{(3)}$ 
and $\overline{\varepsilon}_r$ respectively.
We neglect self-consistent nonlinear effects on the cavity field profile induced
by the Kerr-nonlinearity itself (e.g. field expulsion from the cavity region), which 
could renormalize the effective value of $U_{\mathrm{nl}}$.

The experimental configuration for photon blockade can be modeled by
adding a coherent pumping term to obtain the standard Kerr-type hamiltonian
that is usually employed in quantum optics \cite{drummond80}
\begin{equation}
\hat{H}=\hbar\omega_0 \hat{a}^{\dagger}\hat{a}+
U_{\mathrm{nl}}\hat{a}^{\dagger}\hat{a}^{\dagger}\hat{a}\hat{a}
+F e^{-i\omega_L t} \hat{a}^{\dagger}+F^{\ast} e^{i\omega_L t}\hat{a} \, ,
\end{equation}
where $\Omega=F/\hbar$ is the coherent pumping rate at the laser frequency $\omega_L$.
Losses in the system are quantified either through the intrinsic cavity decay rate, $\kappa$,
or nonlinear absorption processes, such as two-photon absorption (TPA) rate,
$\gamma_{TPA}$. The first is due to coupling of the resonant mode to free space modes, material 
absorption, or scattering from roughness, and defines the cavity quality (Q-) factor
as $Q=\omega_0/ \kappa$; the latter is related to the imaginary part of the nonlinear susceptibility.
Such loss mechanisms are taken into account within a density matrix master equation 
formalism in Markov approximation
\begin{equation}\label{master_eq}
\dot{\rho}=\frac{i}{\hbar}[\rho,\hat{H}] + \mathcal{L}_1(\kappa,\rho)+\mathcal{L}_2(\gamma_{TPA},\rho) \, ,
\end{equation}
where
$\mathcal{L}_1 = \kappa[\hat{a}\rho \hat{a}^{\dagger} - \hat{a}^{\dagger}\hat{a}\rho/2-\rho\hat{a}^{\dagger}\hat{a}/2]$ and
$\mathcal{L}_2 = \gamma_{TPA}[\hat{a}^2\rho (\hat{a}^{\dagger})^2 - (\hat{a}^{\dagger})^2 \hat{a}^2 \rho/2-\rho(\hat{a}^{\dagger})^2 \hat{a}^2/2]$ are
the linear and nonlinear Liouvillian operators, respectively.
In classical nonlinear optics, TPA is quantitatively defined by an intensity-dependent absorption 
coefficient, $\alpha_{TPA}=\beta I$, where $\beta$ is measured in m/W and is well known
for many semiconductor or insulator materials.\cite{boyd_book} Such quantity is related to 
a loss rate, $\gamma_{TPA}=\beta c I /2\overline{n}_r$, where $\overline{n}_r^{2}=\overline{\varepsilon}_r$ 
and $I$ is the field intensity in the cavity.\cite{nota_TPA} 

\begin{figure}[t]
 \begin{center}
  \includegraphics[width=0.48\textwidth]{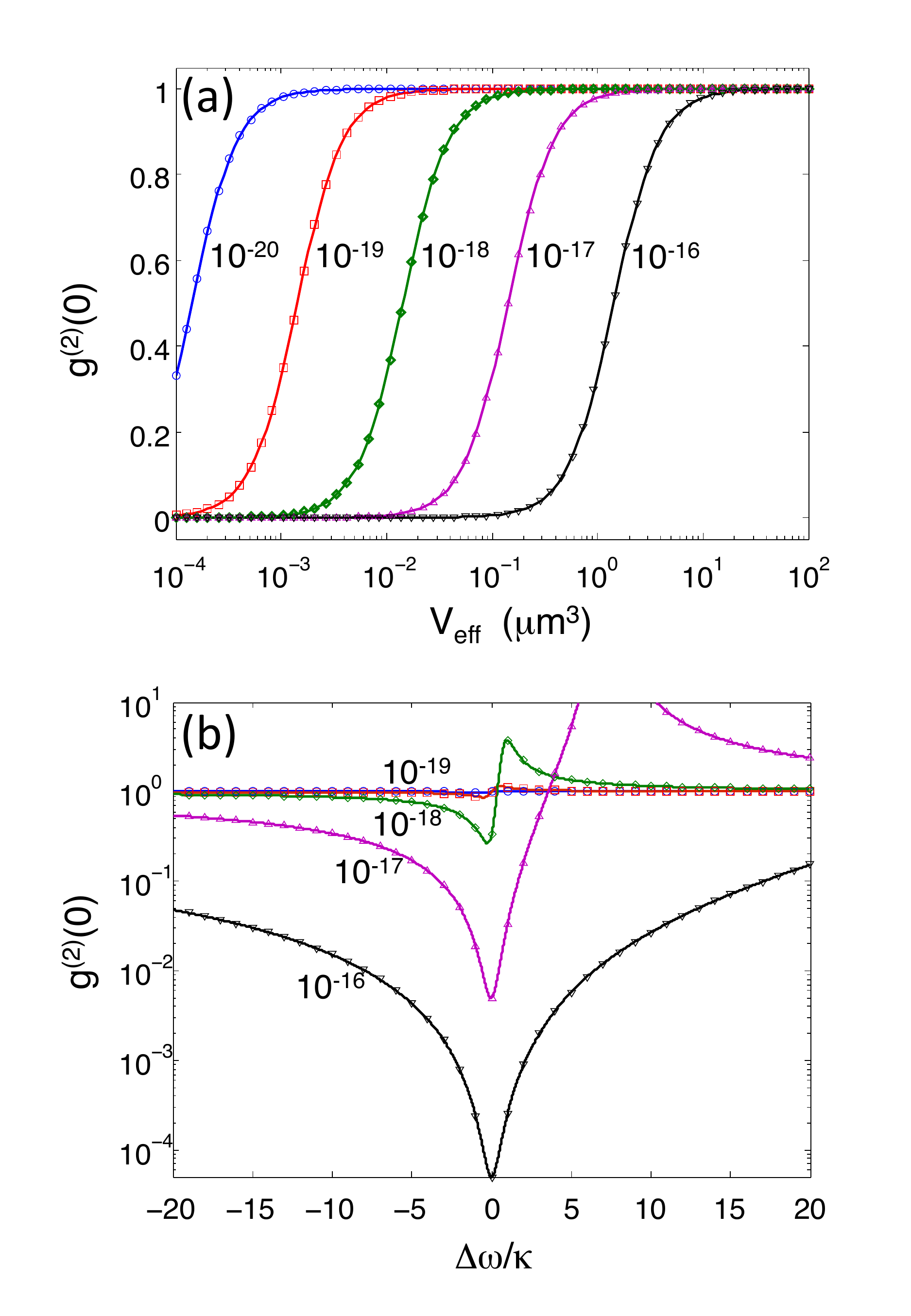}
    \vspace{-1cm} 
   \caption{(Color online) Numerical (symbols) and analytic (full lines) solutions for the zero-time delay 
   second-order correlation for different $\overline{\chi}^{(3)}/ \overline{\varepsilon}^2_r$ values: 
   (a) dependence on effective confinement volume of the cavity field with parameters $\hbar\omega_0=1$ eV, 
   $\Delta \omega=0$, and $Q=10^{6}$, 
   (b) dependence on the detuning between driving field and cavity resonance, for $V_{\mathrm{eff}}=0.01$ 
   $\mu$m$^{3}$ and parameters as before. Numerical results are calculated with $\Omega/ \kappa=0.01$.}
    \label{fig2}
\end{center}
\end{figure}

The figure of merit quantifying the single-photon nonlinear behavior of the cavity mode is
the normalized zero-time delay second-order correlation, defined as 
$g^{(2)}(0) = \langle\hat{a}^{\dag2}\hat{a}^{2}\rangle/ \langle\hat{a}^{\dag}\hat{a}\rangle^2$.
Single photons are released from the cavity at the bare frequency, $\omega_0$. 
In the weak resonant excitation limit ($\Omega/ \kappa \ll 1$)  
a closed analytic solution for the model considered is found after truncating the Hilbert 
space to the $n=2$ Fock state\cite{ferretti2010}
\begin{equation} \label{g2_analytic}
 g^{(2)}(0)=
 \frac{1+4({\Delta E}/{\hbar\kappa})^{2}}{1+4({\Delta E}+{U_{\mathrm{nl}}})^2/{\hbar^2 \kappa}^2}
\end{equation}
where  $\Delta E=\hbar(\omega_L - \omega_0)=\hbar\Delta\omega$.
For $g^{(2)}(0)\to 0$ we have an almost ideal single-photon source,\cite{loudon_book} which occurs when
$U_{\mathrm{nl}}/ (\hbar \kappa)\gg 1$. 
From Eqs.~(\ref{shift_kerr}) and (\ref{g2_analytic}), FOM$=Q^2 / V^{2}_{\mathrm{eff}}$ is the relevant figure of 
merit to be optimized. 

The steady state value of $g^{(2)}(0)$   
can also be calculated numerically through a quantum average on $\rho_{ss}$, 
which is the density matrix corresponding to the eigenvalue $\lambda_{ss}=0$ in the linear 
eigenvalue problem  $\cal{L}\rho = \lambda \rho$.\cite{stenholm03pra} 
Convergence is ensured by truncating the Hilbert space to a large number of photons (up to 50 in this work).
A close agreement between analytic and numerical solutions is reported in Fig.~\ref{fig2}, where 
we show $g^{(2)}(0)$ in the low pumping regime as a function of the confinement volume and 
pump/cavity  detuning, respectively. Results are plotted for different values of the ratio 
$\overline{\chi}^{(3)}/ \overline{\varepsilon}^2_r$, which is a material dependent quantity. 
We assume an operational energy $\hbar\omega_0=1$ eV, 
as representative of typical near-infrared applications, and a 
realistic quality factor $Q=10^6$ (see discussion below).
As shown in Fig.~\ref{fig2}a, for $V_{\mathrm{eff}} \ll \lambda_0^3$ the system exhibits a 
strong antibunching, which is the signature of single-photon blockade. 
For materials with larger $\overline{\chi}^{(3)}/ \overline{\varepsilon}^2_r$ ratios, 
the condition for achieving nonlinear behavior at the single photon level is quantitatively relaxed, 
making it possible to observe strong antibunching even for $V_{\mathrm{eff}} \sim \lambda_0^3$.
Figure~\ref{fig2}a emerges then as a useful roadmap to quantitatively assess, for a specific 
nanostructured material, the combined effect of third-order susceptibility and the confinement 
volume on getting a nonlinear response at the single photon level.
A realistic value $V_{\mathrm{eff}} = 0.01$ $\mu$m$^3$ can be assumed for
diffraction-limited confinement volumes, i.e. $V_{\mathrm{eff}} \sim (\lambda_0/2 \overline{n}_r)^3$,
where $\overline{n}_r$ can display values between 2 and 4, depending on the semiconductor 
or insulator material under investigation.\cite{boyd_book} 
In Fig.~\ref{fig2}b, we show $g^{(2)}(0)$ as a function of pump/cavity detuning and a fixed value
of the confinement volume. 
At large $\overline{\chi}^{(3)}/ \overline{\varepsilon}^2_r$ ratio, the maximum antibunching is 
obtained for $\Delta\omega\sim 0$. At positive detunings, the bunching is  due to the driving laser 
hitting the two-photon resonance of the cavity (see scheme 
in Fig.~\ref{fig1}b).\cite{ciuti06prb}    
Again, in this low pumping regime the quantitative behavior of $g^{(2)}(0)$ obtained from the 
numerical solution is closely reproduced analytically.

\begin{table}[b]
\centering
\begin{tabular}{| c | c | c | c | c | c |}
\hline
Material & $\mathrm{Re}\{\chi^{(3)}\}$ (m$^{2}$/V$^{2}$) &  $\beta$ (m/W) & $n_{r}$ & $\lambda$ ($\mu$m) \\
\hline
  &  &  &  &   \\
Si \cite{dinuAPL03,bristow_apl,hon_jap} & $0.45\times10^{-18}$ & $10^{-11}$ & $3.4$ & $1.55$  \\
Ge \cite{hon_jap} & $4\times10^{-18}$ & $10^{-8}$ & 4.0 & 2.5  \\
GaAs \cite{dinuAPL03} & $0.6\times10^{-18}$ & $10^{-10}$ & 3.4 & 1.54 \\
SiO$_{2}$/Ge \cite{razzari_apl} & $1.4\times10^{-18}$ & $4\times10^{-10}$ & 2 & 0.8 \\
SiO$_{2}$/Si-nc \cite{martinez2010} & $2.1\times10^{-18}$ & $5\times10^{-10}$ & 1.74 & 1.55 \\
SiO$_{2}$/Ag \cite{ganeev_apB} & $7\times10^{-16}$ & $1.5\times10^{-11}$ & 1.8 & 1.06 \\
\hline
\end{tabular}
\caption{Third-order nonlinear optical coefficients of different semiconductor and doped glass materials at
specific wavelengths in the near infrared. }
\label{tab-TSP}
\end{table}

The results shown in Fig.~\ref{fig2} may represent a useful guide to quantum photonics experiments 
employing ordinary nonlinear materials, whose relevant figure of merit for single-photon 
nonlinear behavior can be predicted for any specific nanostructuring-based confinement.
For example, diffraction-limited electromagnetic field confinement can be achieved by using photonic 
crystal nanocavities, in which a number of remarkable figures of merit have been already 
demonstrated experimentally [for a recent review, see Ref. \onlinecite{notomi_review}]. 
Quite interestingly and related to the present work, most of such achievements have been 
obtained by using highly nonlinear materials, such as Silicon (Si) or Gallium Arsenide (GaAs). 
The typical order of magnitude for the $\chi^{(3)}$ tensor elements of these materials
is in the range $\mathrm{Re}\{\chi^{(3)}\}\sim 10^{-19}-10^{-18}$ m$^2 /$V$^2$.\cite{boyd_book} 
However, even larger $\chi^{(3)}$ values can be found in certain 
glasses doped with nanoparticles, chalcogenide glasses, or other polimeric materials.\cite{boyd_book} 
We refer to Table~\ref{tab-TSP} for a few recent experimental references on
the nonlinear coefficients of some Kerr-type materials in the near infrared, which
we have collected from published works and converted in SI units.
Most of these materials can be nanostructured to fabricate solid state nanocavities.
Ultra-high-Q factors in excess of $10^{6}$ have been experimentally
shown, corresponding to a photon lifetime within the cavity region on the order
of one to few ns, and $10^{8}$ has been predicted through design optimization.\cite{notomi_review} 
Designs to achieve sub-diffraction limited mode volumes, on the order of 
$V_{\mathrm{eff}}\sim (\lambda /2n_r)^3$, have been also proposed.\cite{arakawa2011} 
Further reduction of the confinement volume, well below the diffraction
limit, has been predicted for suitably engineered nanostructures.\cite{robinson05prl}
Thanks to these unprecedented figures of merit, strong enhancement of the nonlinear 
optical response due to $\chi^{(3)}$ nonlinearity has been already shown in 
GaAs- and Si-based photonic crystal cavities around $\lambda\sim 1.5$ $\mu$m,
respectively.\cite{matteo_shg}

\begin{figure}[t]
 \begin{center}
  \vspace{-2.8cm}
  \includegraphics[width=0.48\textwidth]{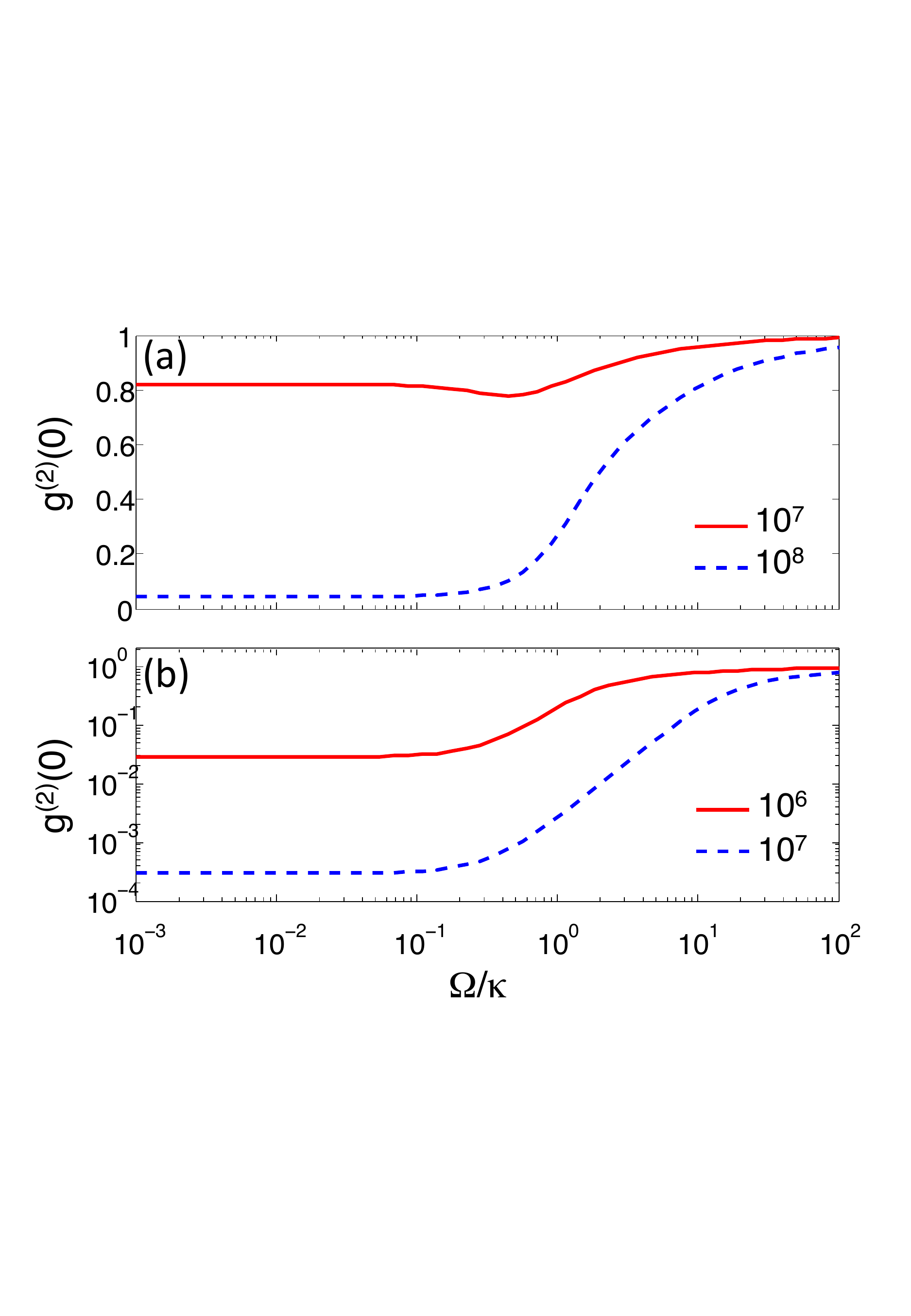}
  \vspace{-3.5cm} 
  \caption{(Color online) Single-photon nonlinear behavior as a function of the driving 
                   strength for a nanocavity made of (a) 
                   a high-index Kerr-type medium with 
                   $\mathrm{Re}\{\chi^{(3)}\}=10^{-18}$ m$^{2} /$V$^{2}$, 
                   $\overline{\varepsilon}_r=10$, $\beta=10^{-10}$ m/W,  
                   $V_{\mathrm{eff}}=10^{-3}$ $\mu$m$^{3}$ ,  or 
                   (b) a low-index, strongly nonlinear material with 
                   $\mathrm{Re}\{\chi^{(3)}\}=10^{-16}$ m$^{2} /$V$^{2}$, 
                   $\overline{\varepsilon}_r=4$, $\beta=10^{-11}$ m/W, $V_{\mathrm{eff}}=10^{-2}$ $\mu$m$^{3}$. 
                   In both cases, $\hbar\omega_0=0.8$ eV ($\lambda_0=1.55$ $\mu$m). 
                   The results shown are: 
                   $Q=10^{7}$ (full line) and $Q=10^{8}$ (dashed line) in (a); 
                   $Q=10^{6}$ (full line) and $Q=10^{7}$ (dashed line) in (b). 
                  }
     \label{fig3}
 \end{center}
 \end{figure}

To quantitatively assess the role of TPA on the $g^{(2)}(0)$ as a function of the pumping strength, 
we have numerically calculated this figure of merit for realistic values of the nonlinear coefficients.
We assume a simple normalized mode profile 
$\alpha(\overline{r})=N\exp(-x^{2}/2\sigma_{x}^{2}-y^{2}/2\sigma_{y}^{2})cos(\pi/d)z$, where the 
normalization factor $N$ is $N=(2/\pi\sigma_{x}\sigma_{y}d)^{1/2}$.
This functional form is a good approximation for a photonic crystal confinement in the $(x,y)$ plane (gaussian
envelope function) and index confinement in the transverse direction, such as the one that can be obtained  with 
a point-defect in a triangular lattice on a planar membrane of thickness $d$.\cite{notomi_review}
From our definition of effective mode volume we have $V_{\mathrm{eff}} = 4 \pi \sigma_x \sigma_y d /3 $.
As illustrative examples, we show results in Fig.~\ref{fig3}  for two different Kerr-type materials.
In Fig.~\ref{fig3}a we assume a high-index ($n_r\sim 3.4$) and strongly nonlinear medium, with 
a typical TPA coefficient in the telecom band.\cite{boyd_book} In such a case, we can assume 
realistic confinement lengths on the order of $\sigma_{x,y}\simeq \lambda_0/ (4 n_r)$ and $d=\lambda_0/(2n_r)$,  
which gives an optimistic $V_{\mathrm{eff}}\simeq 0.001$ $\mu$m$^3$ for wavelengths
on the order of $\lambda_0=1$ $\mu$m.
In Fig.~\ref{fig3}b we assume a low-index ($n_r\sim 2$) material with a sizable Kerr nonlinearity
and negligible TPA coefficient at telecom wavelengths (such as silica with metal nanoparticles, see
table~\ref{tab-TSP}). In such a case, 
the confinement lengths can be $\sigma_{x}\simeq \lambda_0/(2 n_r)$, $\sigma_{y}\simeq \lambda_0/(10 n_r)$,
and $d=\lambda_0(/2n_r)$, where one exploits the slot waveguide confinement at least along one 
direction.\cite{robinson05prl} 
With these numbers at hand, we can again assume $V_{\mathrm{eff}}\simeq 0.01$ $\mu$m$^3$
for this case.  
These results clearly show that efficient single-photon nonlinear behavior can be achieved with 
ordinary Kerr-type media, and that such behavior is robust with respect to nonlinear
sources of dissipation such as TPA. In particular, we notice that in Fig.~\ref{fig3}a TPA contributes 
a nonlinear quality factor $Q_{TPA}\simeq 10^8$ for $\Omega/\kappa\simeq 10^2$, which means
that its effects become relevant only for strong pumping strength and very large  $\omega_0 / \kappa\sim 10^8$.
Realistic Q-factors on the order of $10^7$ can already give clear signatures of single-photon 
nonlinear behavior and sizable antibunching with such high-index media. 
On the other hand, the stronger nonlinearity of doped glasses, together with their negligible TPA effects at
telecom wavelengths, make these materials extremely interesting for quantum photonics applications. 
From Fig.~\ref{fig3}b, Q-factors on the order of $10^6$ are already sufficient
to give an almost ideal single-photon source, provided the confinement volume is as low as the
one assumed. 

So far, we have assumed continuous wave excitation, i.e. $\Omega\neq \Omega(t)$. 
Common solid state single-photon sources exploit the reduced lifetime of a quantum 
emitter in a cavity, allowing single-photon generation on demand at high repetition rates 
through pulsed excitation.\cite{qp_review}
A single-photon source based on the simple scheme of Fig.~\ref{fig1} has the potential advantage 
of working at arbitrary wavelengths (determined by the cavity resonance), with a radiative timescale 
solely determined by the cavity mode characteristic parameters, thanks to the basically instantaneous 
nature of $\chi^{(3)}$  processes.\cite{boyd_book}  
Thus, in a pulsed excitation scheme the requirements on the resonant laser 
source are determined by the constraints on the pulse duration ($\hbar/U_{\mathrm{nl}} < \Delta t < \kappa^{-1}$) 
and period ($\Delta T \geq 5\kappa^{-1}$) preserving photon blockade.\cite{imamoglu97,ciuti06prb} 
From the results shown in Fig.~\ref{fig2}b, the device would also be tolerant to possible fluctuations of 
the laser center frequency around the cavity resonance, which are normally smaller than
the cavity linewidth in standard near-infrared laser sources. 
With a $Q\simeq 10^6$, i.e. $\kappa^{-1}\sim 1$ ns, and  $U_{\mathrm{nl}}$ of a few $\mu$eV, the pulse 
duration should be between 0.1 and 1 ns, while the maximum repetition rate would be limited to a few hundred MHz,
which is comparable to the fastest single-photon source on demand recently demonstrated with solid 
state quantum emitters.\cite{rivoire_sps}
The potential repetition rate can be further increased by relaxing the requirements on the Q-factor, 
i.e. by increasing $U_{\mathrm{nl}}$ through reduction of the cavity mode volume, anticipating much more 
controllability and flexibility as compared to single quantum emitters.

In summary, we have shown that future quantum photonics applications can strongly benefit from 
the capability of nanostructuring ordinary Kerr-type materials to achieve sub-diffraction limited 
electromagnetic field confinement.
The growing interest in integrated quantum photonics,\cite{politi08sci} and the possibility of 
fully exploiting the mature CMOS-based technology to build room-temperature and intrinsically
flexible single-photon devices are likely to produce new research avenues based on 
the present proposal in the near future.

\vspace{0.1cm}

\begin{acknowledgments}
We ackowledge useful and several motivating discussions  with 
L.C. Andreani, D. Bajoni, I. Carusotto, M. Galli, M. Liscidini, L. O'Faolain, V. Savona,  H. T\"{u}reci. 
\end{acknowledgments}


\begin{thebibliography}{99}
\bibitem{nota_TPA}
The absorption coefficient is related to the imaginary part 
of frequency (and hence to the loss rate) as $\alpha_{TPA}=2\mathrm{Im}(\omega) n_r /c$, 
where $n_r$ is the material refractive index and $c$ is the speed of light.
\end{thebibliography}
\end{document}